\documentclass[useAMS,usenatbib]{mn2e}

\usepackage{times}

\usepackage{graphicx}
\usepackage{times}
\newcommand{\cthead}[1]{\multicolumn{1}{c}{#1}}
\newcommand{\kss}{km~s$^{-1}$ }
\newcommand{\ks}{km~s$^{-1}$}
\title[New class~I methanol maser at 23.4~GHz]{Discovery of the new class~I methanol maser transition at 23.4~GHz}
\author[M. A. Voronkov et al.]{M. A. Voronkov$^{1,2}$\thanks{E-mail:
Maxim.Voronkov@csiro.au}, A. J. Walsh$^{3}$, J. L. Caswell$^1$, 
S. P. Ellingsen$^{4}$, S. L. Breen$^{1,4}$, \newauthor 
S. N. Longmore$^{5}$, C. R. Purcell$^{6}$,  J. S. Urquhart$^{1}$\\
$^{1}$Australia Telescope National Facility, CSIRO Astronomy and Space Science, PO Box 76, Epping,
NSW 1710, Australia\\
$^{2}$Astro Space Centre, Profsouznaya st. 84/32, 117997 Moscow, Russia\\
$^{3}$Centre for Astronomy, School of Engineering and Physical Sciences, James Cook University, Townsville, QLD 4814, Australia\\
$^{4}$School of Mathematics and Physics, University of Tasmania, GPO Box
252-37, Hobart, Tasmania 7000, Australia\\
$^{5}$ESO Headquarters, Karl-Schwarzschild-Str, 2, 85748, Garching bei M\"unchen, Germany\\
$^{6}$School of Physics and Astronomy, University of Leeds, Leeds, LS2 9JT, UK\\}
\begin{document}

\date{}

\pagerange{\pageref{firstpage}--\pageref{lastpage}} \pubyear{2010}

\maketitle

\label{firstpage}

\begin{abstract}
We report the first detection of a methanol maser in the $10_1-9_2$~A$^-$ transition at
23.4~GHz, discovered during the H$_2$O southern Galactic Plane Survey (HOPS) with 
the 22-m Mopra radio telescope. In the region covered by HOPS, the 23.4-GHz maser was 
found at only one location, G357.97$-$0.16, which was also a prominent source of 
maser emission in the J$_2-$J$_1$~E series near 25-GHz. The Australia Telescope 
Compact Array (ATCA) was used to follow up these detections at high angular resolution and
prove the maser nature of the observed emission.  The analysis shows that the new methanol 
maser at 23.4-GHz is a class~I maser, which has properties similar to the 9.9 and 25-GHz 
masers (i.e. traces strong shocks with higher than average temperature and density). 
All class~I masers were found to originate at the same spatial location (within the measurement 
uncertainty of 0.5 arcseconds) in the vicinity of the dominant infrared source, but at a clearly 
distinct position from
nearby OH, H$_2$O and class~II methanol masers at 6.7-GHz. All maser species are 
distributed approximately on a line, but it is not clear at present whether this has any physical 
significance. We also detected a weak (1.3~mJy) continuum source at 25-GHz near 
the OH maser (at the most northern site, associated with a class~II methanol maser and 
an H$_2$O maser renowned for its extremely wide spread of velocity components). 
The continuum source has not been
reported at lower frequencies and is therefore a candidate hypercompact H{\sc ii} region.
We also used the ATCA to find the strongest
and only fifth known 9.9-GHz maser towards G357.97$-$0.16 and another 23.4-GHz maser
towards G343.12$-$0.06 not seen in HOPS.
\end{abstract}


\section{Introduction}

Methanol masers are associated with regions of active star formation,
with more than twenty different centimetre and millimetre wavelength
maser transitions discovered to date \citep[e.g.,][]{mul04}. There are 
two distinct classes of methanol masers as suggested empirically in 
the early studies \citep{bat87}. Widespread class~I masers such as
44 and 95~GHz masers usually occur in multiple locations across the
star-forming region scattered around an area up to a parsec in extent
 \citep[e.g.][]{kur04,vor06,cyg09}. In contrast, class~II methanol masers
 (e.g. at 6.7 and 12~GHz), along with OH and H$_2$O masers, reside in the close 
 vicinity of exciting young stellar objects (YSOs) and are typically found as a 
 single cluster of emission at arcsecond resolution \citep[e.g.,][]{cas10}.
 
Theoretical calculations can explain 
this empirical classification and strongly suggest that the pumping process of 
class~I masers is dominated by collisions with molecular hydrogen, in contrast to 
class~II masers which are pumped by radiative excitation \citep[e.g.][]{cra92}. 
The two pumping mechanisms were shown to be competitive 
\citep[for an illustration, see][]{vor05}: 
strong radiation from a nearby infrared source quenches class~I masers and 
increases the strength of  class~II masers.  Therefore, bright masers of different classes 
residing in the same volume of gas are widely accepted as mutually exclusive (note that there
could be exceptions for weak masers).  However, on larger scales, they are often observed to 
coexist in the same star forming region within less than a parsec of each other. The exceptions
are a few archetypal sources where only one particular class of methanol masers has been 
detected. 

The class~I methanol masers are less studied than the class~II counterparts but have recently 
become a subject of intense research. The common consensus is that class~I masers trace 
shocked gas, where the conditions favour collisional excitation and a significant amount of
methanol is released from dust grain mantles.  The exact physical phenomena causing 
shocks traced by masers probably vary from source to source 
\citep[see, e.g.,][and references therein]{vor10}, 
although for most of these masers, the shocks probably arise from an interaction between outflows 
and the ambient molecular material \citep[e.g.][]{pla90,kur04,vor06,cyg09}. 

In addition to the gross classification, there are finer distinctions within the same class of methanol
maser transitions. At sub-Jy sensitivity levels, the range of transitions
can be further categorised into widespread masers and rare or weak masers. For example, more than 
100 maser sources are known at 44~GHz \citep[e.g.][]{has90,sly94,cyg09}, while very few were found 
at 9.9~GHz despite a sensitive search towards a large number of targets \citep{vor10}. 
Another example of rarer masers is the J$_2-$J$_1$~E series near 25~GHz.
Historically, these were the first methanol masers found in space \citep{bar71,hil75}, 
but were believed to be rare following the survey of \citet{men86}. However, it has recently been
shown by a more sensitive survey that the 25~GHz masers are quite common, but weak 
\citep[typically weaker than 1~Jy;][]{vor07}. 
The modelling to date suggests that class~I masers at 9.9~GHz and the series of masers near 
25-GHz require higher temperatures and densities to form than their widespread counterparts
\citep{sob05}. This is in agreement with the less frequent appearance of these masers. 
In this paper we present the first detection of a new class~I methanol maser at  23.4-GHz, which
shows similar properties to 9.9 and 25-GHz masers and may shed light on class~I methanol maser
origins.


\section{Observations}

\begin{table}
 \caption{Methanol maser transitions observed with ATCA.}
 \label{obsdetails}
 \begin{tabular}{@{}lc@{}r@{}c@{}r}
\hline
 \cthead{Molecular}  & \cthead{Rest} & \multicolumn{2}{c}{Rest frequency} 
 & \cthead{Velocity}  \\ 
 \cthead{transition}& \cthead{frequency} &\multicolumn{2}{c}{uncertainty} &\cthead{range} \\
&  \cthead{(MHz)} &\cthead{(MHz)}&\cthead{(\ks)}&\cthead{(\ks)}  \\
  \hline
\multicolumn{5}{c}{19 June 2010} \\
$\hphantom{1}2_2-2_1$~E & 24934.382\hphantom{0} & 0.005\hphantom{0} & 0.060 & $-$9.3, \hphantom{$-$}0.8 \\
$\hphantom{1}3_2-3_1$~E & 24928.707\hphantom{0} & 0.007\hphantom{0} & 0.084 & $-$11.5, $-$1.2 \\
$\hphantom{1}4_2-4_1$~E &  24933.468\hphantom{0} & 0.002\hphantom{0} & 0.024& $-$8.3, \hphantom{$-$}2.0 \\
$\hphantom{1}5_2-5_1$~E & 24959.0789 & 0.0004 & 0.005 & $-$6.9, \hphantom{$-$}3.3 \\
$\hphantom{1}6_2-6_1$~E &  25018.1225 & 0.0004 & 0.005 & $-$12.2, $-$2.1 \\
$\hphantom{1}7_2-7_1$~E &  25124.8719 & 0.0004 & 0.005 & $-$9.4, \hphantom{$-$}0.9 \\
$\hphantom{1}8_2-8_1$~E &  25294.4165 & 0.0004 & 0.005 & $-$8.8, \hphantom{$-$}1.5 \\
$10_1-9_2$~A$^-$ & 23444.778\hphantom{0} & 0.002\hphantom{0} & 0.026 & $-$11.9, $-$0.9 \\
\multicolumn{5}{c}{ 7 June 2010}\\
$9_{-1}-8_{-2}$~E & 9936.201\hphantom{8} & 0.001\hphantom{0} & 0.030 & $-$22.7, \hphantom{$-$}4.5 \\
\hline
\end{tabular}
\end{table}

The initial detection of the $10_1-9_2$~A$^-$ methanol maser at 23.4~GHz was made as
part of the H$_2$O southern Galactic Plane Survey \citep[HOPS; see also][]{wal08} towards
a single location G357.97$-$0.16. HOPS is an unbiased survey of the southern Galactic plane 
at frequencies between 19.5 and 27.5~GHz, carried out using the Mopra radio telescope. 
It is designed primarily to detect water masers and NH$_3$ emission, but also covered a 
range of methanol maser transitions of both class~I and class~II.
The zoom mode of the Mopra spectrometer (MOPS) was used in the survey. It provided
16 simultaneous spectral windows, each 137.5~MHz wide and split into 4096 channels, 
positioned within the 8.3-GHz wide receiver band around the frequencies of interest. The 
attained spectral resolution varied with frequency from 0.3~\kss to 0.5~\ks. HOPS has
covered 100 square degrees of the Galactic plane between the longitudes 290\degr and 
30\degr{~}through the Galactic centre and latitudes  $\pm$0.5\degr{~}with the typical
1-$\sigma$ noise level of around 1$-$2~Jy. It is worth noting, however, that a high signal-to-noise
ratio is required to detect masers (at least $5\sigma$, if not more) because the spectral resolution of 
HOPS is relatively coarse and the masers are expected to appear as single channel spikes.  
Further details on the survey will be published in a technique paper (Walsh et al., in prep.).

The HOPS detection of the 23.4~GHz maser and the J$_2-$J$_1$~E maser series near 25~GHz 
towards G357.97$-$0.16 was followed up with the Australia Telescope Compact Array (ATCA)
on 19 June 2010 (project code CX110) using recently commissioned zoom modes of the Compact Array Broadband Backend (CABB), which allowed us to observe 8 maser transitions simultaneously. 
In addition to this, we searched for another rare methanol maser at 9.9~GHz towards this source
in a separate observation on June 7th.  The full list of observed transitions is 
given in Table~\ref{obsdetails}
along with the rest frequency for each transition, corresponding velocity uncertainty and 
observed velocity range. The majority of rest frequencies were taken from 
\citet{mul04}. However, we adopted an astronomical measurement of the rest frequency 
for the 9.9-GHz transition \citep{vor06}. It agrees with the laboratory measurement of 
\citet{mul04}, which has the uncertainty of 0.12~\ks. The rest frequency for 
the 23.4~GHz transition was taken from \citet{meh85}.

The correlator split each 1-MHz wide zoom window into 
2048 spectral channels providing a spectral resolution of about  0.015~\kss and 0.006~\kss  for
the 9.9 and 23-25-GHz observing sessions, respectively. In addition, two 2-GHz wide windows 
with 1~MHz spectral resolution were also available for simultaneous continuum measurement. 
These broadband windows were set to overlap in order to accommodate the particular 
configuration of zoom windows used for the spectral line observations. This set up 
provided an effective bandwidth of 3~GHz (the overlap region does not improve signal-to-noise ratio).

The ATCA antennas were in 
an extended 6C array configuration for both observing slots giving baselines ranging 
from 153~m up to 6~km\footnote{More details on the ATCA
configurations are available on the web ({\it http://www.narrabri.atnf.csiro.au/observing/configs.html})}.
Note, however, that CA02 antenna forming the shortest baseline of the 6C configuration 
was not available during the 23-25-GHz observations on June 19th (the second shortest baseline 
was 413~m).  Both observing slots were short (6 hours on June 19 and just 30 minutes on June 7)  resulting in an asymmetric synthesised beam of about $1.3\times0.3$ arcseconds at
position angle of $-$38\degr{ }for the 23-25~GHz transitions and precluding any imaging at 9.9~GHz. 
For the latter, we performed phase-only self-calibration with a solution interval of 3 minutes and
assumed a point source model at the position of the strongest component of the 23.4~GHz maser 
(see the following section) to extract the spectrum. 

The position of the phase and pointing centre was
$\alpha_{2000}=17^h41^m$19\fs12, $\delta_{2000}=-30$\degr44\arcmin58\farcs06 for both
observing slots. For the June 19th slot we used reference pointing procedures and 
determined corrections using  the continuum source 1714-336 (which served also as a phase calibrator). From the statistics of the pointing solutions the reference pointing accuracy was 
estimated to be 2\farcs6$\pm$2\farcs2.  The 9.9-GHz observations were
done using the global pointing model which is expected to be accurate to about 10~arcsec.
The primary beam size was about 2 arcmin at 23-25~GHz and 5.1~arcmin at 9.9~GHz. 
Note that pointing errors affect the accuracy of flux density measurements, while the accuracy of 
the obtained absolute positions depends primarily on the quality of 
the phase calibration and is believed to be better than 0.5~arcsec. At the 
position of the masers, which are discussed in the following section, these pointing uncertainties
correspond to flux density uncertainties of 1\% and 2\% at 9.9 and 23-25~GHz, respectively.
The absolute flux density scale was bootstrapped from observations of 1934-638. We expect it
to be accurate to better than 3\% and 10\% at 9.9 and 23-25~GHz, respectively.  The assumed 
flux densities were 2.39 and 0.75~Jy at 9.9 and 23.4~GHz, respectively.

The data reduction was performed using  the {\sc miriad} package (CABB release) following standard
procedures and ignoring wide-bandwidth effects for the continuum measurement (these
effects are negligible for narrow zoom windows). Each spectral 
window produced by CABB was processed  independently, although we merged the overlapping broadband windows together for imaging after the 
calibration and removal of contaminating spectral lines. The imaging was performed for the whole 
field of view and used natural weighting.
We searched for maser emission in the image cube prior to the primary beam correction (i.e. the noise 
across the field of view was constant). Then, the cube was divided by the primary beam model at 
the appropriate frequency and the spectra were extracted at the peak pixel by taking a slice along 
the spectral axis. We followed this approach due to the rather high sidelobe level in the point spread function caused by the poor uv-coverage attained in the project. This method reproduces the flux density
correctly in the case of unresolved or barely resolved sources  (i.e. smaller  than the synthesised beam) 
and is well suited to maser observations.

\section{Results}
\subsection{ATCA spectra of the masers}

\begin{figure}
\includegraphics[width=\linewidth]{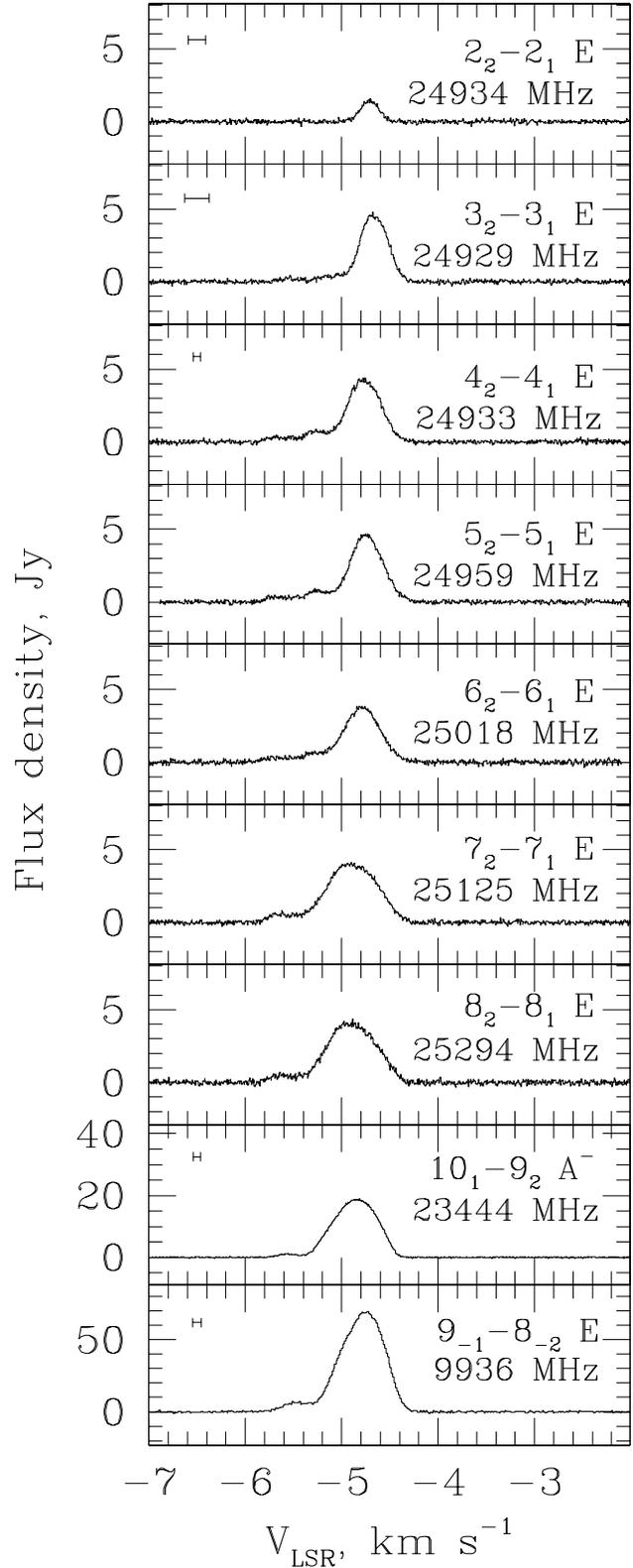}
\caption{Spectra of the masers detected in G357.97$-$0.16. Horizontal error bars
show the 3$\sigma$ uncertainty in radial velocity caused by the rest frequency uncertainty
(the error bars are only shown if the uncertainty notably exceeds the spectral resolution).}
\label{spectra}
\end{figure}

We detected emission in all class~I maser transitions observed towards G357.97$-$0.16 
(see Table~\ref{obsdetails}). This is an independent confirmation of the HOPS detections. 
In addition, the search for the 9.9-GHz maser yielded a detection of only the fifth maser
at this frequency. It has a peak flux density of around 70~Jy, exceeding that of all other known 9.9-GHz
masers by an order of magnitude \citep[c.f.][]{vor10}. 
The spectra of all observed transitions are shown in Figure~\ref{spectra}.
It is clear that a number of spectral components contribute to the overall profile for each 
transition. 
The relative flux densities of these components vary with transition which is reflected by the 
shapes of the spectra in Figure~\ref{spectra} and by the slightly broader appearance of the spectral 
profiles corresponding to the 
high excitation J=7 and 8 transitions of the J$_2-$J$_1$~E series in comparison to
the lower J transitions. The apparent offset in velocity (most pronounced for the J=3 transition) is
consistent with the rest frequency uncertainty. 

\begin{table*}
\caption{Fit results and profile parameters. The uncertainties are
given in parentheses and expressed in units of the least significant
figure.}
\label{fit_results}
\begin{tabular}{@{}lllrrrlrr}
\hline
 & \multicolumn{5}{c}{Gaussian components} & \cthead{Peak} &
\cthead{Peak} \\
\cthead{Molecular}  & \cthead{LSR} & \cthead{$\alpha_{2000}$} & \cthead{$\delta_{2000}$}&
\cthead{Line} & \cthead{Flux} &  \cthead{LSR} &
\cthead{flux} &\cthead{$\int f(v)\;dv$}\\
 \cthead{transition}  & \cthead{Velocity\makebox[0mm]{\hskip 2mm $^a$}} & \cthead{17$^h$41$^m$} & \cthead{$-$30\degr45\arcmin} &
\cthead{FWHM} & \cthead{density} & 
\cthead{velocity\makebox[0mm]{\hskip 2mm $^b$}} &
\cthead{density} &\\
& \cthead{(\ks)} & \cthead{($^s$)}&\cthead{(arcsec)} &
\cthead{(\ks)} & \cthead{(Jy)} &  \cthead{(\ks)} &
\cthead{(Jy)}&\cthead{(Jy \ks)}  \\
\hline
$\hphantom{1}2_2-2_1$~E &  $-$4.710 & 20.051~(6) & 18.1\hphantom{1}~(1) &   0.224~(4) & 1.43~(2) & $-$4.703 & 1.6~(2) & 0.34~(3) \\
$\hphantom{1}3_2-3_1$~E & $-$5.53\hphantom{1} &  20.056~(8) & 18.1\hphantom{1}~(1)  & 0.25\hphantom{7}~(3) & 0.22~(2) & $-$4.679 & 4.8~(5) & 1.8\hphantom{4}~(2) \\
  &  $-$5.14\hphantom{1} & 20.067~(9) & 18.4\hphantom{1}~(2) & 0.25\hphantom{1}~(2) & 0.39~(2) \\
  &  $-$4.665 & 20.049~(5) & 18.06~(7) & 0.333~(2) & 4.67~(2) \\
$\hphantom{1}4_2-4_1$~E  & $-$5.64\hphantom{2} & 20.04\hphantom{1}~(2) & 17.9\hphantom{1}~(3)  & 0.30\hphantom{4}~(3) & 0.32~(2) & $-$4.756 & 4.4~(5) & 2.1\hphantom{3}~(2)\\
 & $-$5.280 & 20.07\hphantom{1}~(2) & 18.5\hphantom{1}~(2)  & 0.23\hphantom{7}~(1) & 0.70~(2) \\
 & $-$4.766 & 20.050~(4) & 18.07~(6) & 0.399~(2) & 4.32~(2) \\
 $\hphantom{1}5_2-5_1$~E  &$-$5.63\hphantom{7} & 20.05\hphantom{1}~(1) & 18.03~(2) & 0.33\hphantom{2}~(3) & 0.37~(2) & $-$4.733 & 4.7~(5) & 2.2\hphantom{3}~(2) \\ 
 & $-$5.265 & 20.064~(9) & 18.3\hphantom{3}~(2) & 0.24\hphantom{2}~(1) & 0.74~(2) \\
 & $-$4.744 & 20.051~(5) & 18.10~(7)& 0.392~(2) & 4.60~(2) \\
 $\hphantom{1}6_2-6_1$~E & $-$5.67\hphantom{0} & 20.05\hphantom{1}~(2)& 18.3\hphantom{1}~(3) & 0.34\hphantom{6}~(4) & 0.31~(2) & $-$4.757 & 3.8~(4)& 2.0\hphantom{3}~(2) \\
  & $-$5.302 & 20.06\hphantom{3}~(1) & 18.3\hphantom{1}~(1) & 0.27\hphantom{2}~(2) & 0.53~(2) \\
  & $-$4.788 & 20.050~(6) & 18.05~(7) & 0.433~(3) & 3.78~(2) \\
 $\hphantom{1}7_2-7_1$~E & $-$5.63\hphantom{1} & 20.06~(1) & 18.2\hphantom{6}~(2) & 0.30\hphantom{1}~(2) & 0.56~(2) & $-$4.901 & 4.1~(4) & 2.7\hphantom{3}~(3)\\
  & $-$4.95\hphantom{1} & 20.050~(4) & 18.06~(7) & 0.50\hphantom{1}~(1) & 3.87~(8) \\
  & $-$4.64\hphantom{1} & 20.048~(6) & 18.02~(9) & 0.33\hphantom{1}~(2) & 1.4\hphantom{1}~(2) \\
 $\hphantom{1}8_2-8_1$~E  & $-$5.625 & 20.059~(8) & 18.2\hphantom{1}~(2) & 0.28\hphantom{1}~(2) & 0.49~(3) & $-$4.886 & 4.3~(4) & 2.6\hphantom{3}~(3)\\
  &  $-$4.96\hphantom{1} & 20.051~(5) & 18.09~(7) & 0.48\hphantom{1}~(2) & 3.94~(8) \\
  &  $-$4.64\hphantom{4} &  20.051~(8) & 18.1\hphantom{7}~(1) & 0.33\hphantom{1}~(2) & 1.5\hphantom{1}~(2)\\
$10_1-9_2$~A$^-$ &  $-$5.584 & 20.049~(8) & 18.0\hphantom{7}~(1) & 0.191~(8) & 1.00~(4) & $-$4.814 & 19\hphantom{.1}~(2) & 11\hphantom{.73}~(1)\\
   & $-$4.932 & 20.050~(3) & 18.07~(3) & 0.450~(5) & 16.4\hphantom{1}~(2) \\
   & $-$4.675 & 20.050~(3) & 18.07~(5) & 0.301~(4) & 8.6\hphantom{1}~(4) \\
$9_{-1}-8_{-2}$~E & $-$5.493 & \multicolumn{2}{c}{no imaging} & 0.30\hphantom{0}~(1) & 6.1\hphantom{8}~(1) & $-$4.73\hphantom{1} & 69\hphantom{.1}~(2) & 40\hphantom{.83}~(1)\\   
& $-$4.89\hphantom{6} & & & 0.44\hphantom{0}~(2) & 50\hphantom{.40}~(2) \\
& $-$4.638 & & & 0.34\hphantom{5}~(7) & 41\hphantom{.98}~(4) \\
\hline
\end{tabular}
\begin{flushleft}
\par\noindent
$^a$The fit uncertainty is half of that for the line FWHM.
\par\noindent
$^b$The uncertainty is the spectral resolution.
\end{flushleft}
\end{table*}

To analyse the  profiles in detail we decomposed
the spectra into a number of Gaussian components listed in Table~\ref{fit_results}. The first column
shows the molecular transition in the same order as in Figure~\ref{spectra}. The following five
columns represent the peak velocity, absolute position (except for the 9.9-GHz components, 
for which no position measurement was possible), line width given as the full width at half 
maximum (FWHM) and the flux density for every Gaussian component constituting the profile.
Uncertainties are given in brackets and expressed in the units of the least significant figure.
They are the formal uncertainties of the fit and do not take into account systematic effects.
It is evident from Table~\ref{fit_results} that all components of all measured transitions arise
at the same location within the positional uncertainty of our measurement (about  0.5 arcsec). 
Therefore, the 9.9-GHz 
profile, which was obtained towards the same position by choice, correctly represents the 
strength of the different components contributing to the profile. We used the smallest number of
Gaussian components required to fit the profile within the measurement errors. However, it
is evident from both Table~\ref{fit_results} and Figure~\ref{spectra} that the radial velocities 
of components corresponding to the J$\le$6 transitions of the 25-GHz series are different 
from that of the rest of the transitions which correspond to higher excitation energies. It is 
therefore likely that more than three physical components are present, but some fade away in
transitions with higher excitation energies while others become brighter.
The last three columns describe each measured spectral profile as a whole prior to decomposition into
components. These columns are the peak velocity, the corresponding flux density and
the integrated flux density over the line profile. The corresponding uncertainties of the 
flux density and the integral take into account the accuracy of the absolute flux 
scale calibration.

Astronomical observations are sometimes used to improve the accuracy to 
which the transition rest frequencies are known \citep[e.g.,][]{vor06}.  The main
drawback of this method is the assumption of equal radial velocities for different
transitions, which may be misleading due to unresolved source structure and
varying excitation conditions. However, it is evident from Figure~\ref{spectra} and 
Table~\ref{fit_results} that the spectral profiles for the J$<$6 transitions of the 
25-GHz series look very consistent. Therefore, we can refine the rest
frequencies of the lower J transitions using the J=5 transition as a reference. The latter has
the rest frequency known to higher accuracy than the spectral resolution of our 
measurement (see Table~\ref{obsdetails}). Assuming that the brightest Gaussian 
component from Table~\ref{fit_results} peaks at the same velocity, the rest 
frequencies are 24934.379$\pm$0.001,
24928.7004$\pm$0.0008 and 24933.4698$\pm$0.0008~MHz, for the J=2, 3 and 4
transitions of the J$_2-$J$_1$~E methanol series, respectively. The uncertainties
are determined by the accuracy of the fit combined with the uncertainty of the rest frequency 
for the J=5 transition and the spectral resolution. It is not practical to adjust the rest frequency 
for other transitions.

\subsection{Morphology of the region}

\begin{figure}
\includegraphics[width=\linewidth]{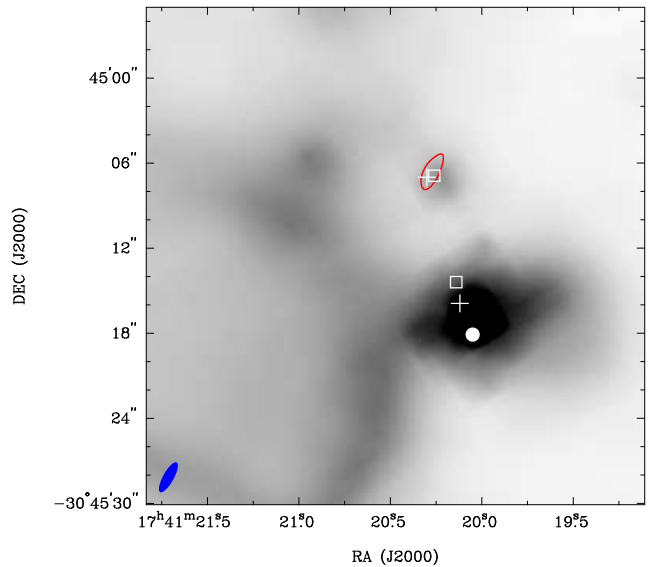}
\caption{Position of class~I methanol masers (circle), class~II methanol masers
at 6.7-GHz (squares) and water masers (crosses) in G357.97$-$0.16 overlaid on top
of the 8.0~$\mu$m IRAC {\it Spitzer} image. The 12-mm continuum source, which has the peak
flux density of 1.3~mJy at 25~GHz, is shown by a single 50\% contour. The ellipse in the bottom
left corner represents the synthesised beam of the continuum measurement.}
\label{masmap}
\end{figure}

The region is a known site of water, OH and class~II methanol masers at 6.7~GHz
\citep{tay93,bre10,cas98,cas10}. The morphology of the region is shown in Figure~\ref{masmap}, 
where the positions of various masers are overlaid on the 8.0-$\mu$m image obtained
with the {\it Spitzer Space Telescope's} Infrared Array Camera (IRAC). Inspection of other IRAC
bands revealed no  pronounced diffuse emission  at 3.6, 4.5 and 5.8~$\mu$m similar to the 
diffuse structures seen at 8.0~$\mu$m
(see Fig.~\ref{masmap}). The dominant infrared source located near the position of class~I
masers is likely to be saturated in all IRAC bands and therefore the emission 
excess at 4.5~$\mu$m, which is often regarded as a signature of shocked 
gas \citep[see, e.g.,][]{deb10}, cannot be determined. Likewise, no Extended Green Object 
(EGO; extended emission with an excess of flux density at 4.5~$\mu$m) has been 
reported in this region.

There are two distinct sites of water and 6.7-GHz methanol maser emission separated by about
8 seconds of arc (see Figure~\ref{masmap}). The southern site is located close to the dominant
infrared source, while the northern maser site is located near a weak source
seen only at 8.0~$\mu$m, the longest IRAC wavelength.  This source is detected in 24-$\mu$m
{\it Spitzer} observations, which means that it is most likely deeply embedded in the parent molecular cloud. 
The hydroxyl maser has been detected towards the
northern site only \citep[located within 2 seconds of arc from other masers;][]{cas98}. This OH
maser has a velocity range from $-$10 to $-$4~\kss \citep{cas98}, while the 
6.7-GHz methanol emission in the northern site is confined to radial velocities 
from $-$6 to 0~\ks.  In contrast, the water
maser in the northern site has an unusually wide  spread of velocity components
\citep[about 180~\protect\kss wide;][]{bre10} and near continuous emission across the 
whole velocity range. 

The companion water maser in the southern site shows much more modest velocity spread
of about 15~\kss \citep{bre10}. The corresponding 6.7-GHz methanol maser is notably weaker
than the maser in the northern site \citep[approximately 3~Jy versus 50~Jy in the northern site;][]{cas10},
but has a slightly wider velocity spread from $-$9 to $+$3~\ks.  
The class~I methanol masers are located a 
few arcseconds offset from the southern site of water and 
class~II methanol masers (near the opposite edge of the brightest infrared peak, 
see Figure~\ref{masmap}). Their radial velocities, which are 
typically found close to the systemic velocity \citep[see e.g.,][]{vor06}, are near $-$5~\kss
(see Table~\ref{fit_results}) and  fall within the velocity interval of the class~II
maser. Despite the fact that this source is observed in the general direction of the 
Galactic Centre, the kinematics do not allow us to make a definite association. 
With the systemic velocity around $-$5~\ks, the distance to this source is 
highly uncertain \citep[see also the discussion in][]{cas10}.

It is worth noting, that the class~I masers are located
approximately on the extension of the line joining the northern and southern
maser sites. However, the available data on the source
do not allow us to establish whether these sites of maser emission are somehow physically
related (e.g. trace a common outflow). It is also not clear whether there is a 
connection between an unusual water maser in the northern site and the rare class~I methanol 
maser in the southern site. Interferometric observations of the widespread class~I methanol
masers (such as masers at 36 and 44~GHz) towards this source are likely to shed light on
this question. These masers often consist of numerous spots aligned in a regular structure, which
presumably traces shocks, and, therefore, have a high potential to reveal an outflow
\citep[e.g.,][]{kur04,vor06,cyg09}. 

The complementary continuum measurement yielded detection of a 
weak (peak flux density is 1.3$\pm$0.4~mJy~beam$^{-1}$ at 25~GHz) compact source
near the northern group of masers (Fig.~\ref{masmap}). 
The present data do not allow us to estimate the spectral index with sufficient accuracy to be able
to constrain emission models.
The source appears to be very slightly resolved, although observations with a better uv-coverage
are required to confirm this. This is the first detection of a radio continuum
source towards this position.  The source is more than an order of magnitude weaker than the 
continuum detection limit of \citet{bre10}. It remained undetected in the 5-GHz survey of 
\citet{bec94} with a flux density limit of around 2.5~mJy. Therefore, this continuum source is 
a good candidate to be a hypercompact H{\sc ii} region, although the present low frequency 
flux density limit does not allow us to rule out other possibilities like a jet, stellar wind or an 
ultra-compact H{\sc ii} region.

\section{Discussion}


The $10_1-9_2$~A$^-$ methanol transition at 23.4~GHz was first observed in absorption 
towards an archetypal class~II methanol maser source W3(OH) by \citet{men85}. This fact
alone suggests that the maser in this transition should belong to class~I. This is 
because the maser pumping influences level populations  in the opposite way for the 
strong maser transitions of different classes 
\citep[for details see Fig.~5 and the associated discussion in][]{vor05}, which can sometimes lead 
to absorption. It is also worth noting, that \citet{sob07} conducted sensitive observations of
20 targets at 23.4~GHz as part of their search for the 23.1-GHz class~II methanol masers.
No emission at 23.4~GHz was detected towards the sources where the class~II maser was also found.
The attribution of the 23.4-GHz masers to class~I is 
corroborated by the pumping models of \citet{cra92}, which predicted these masers 
when collisional transitions dominate (i.e. the case of class~I masers).
The methanol emission in G357.97$-$0.16 was found to be unresolved for all observed 
transitions, despite using the most extended array configuration of ATCA. This implies the brightness
temperature is in excess of 10$^6$~K and confirms the maser nature of the observed emission.
This conclusion is further reinforced by the narrow width of the spectral profiles 
(see Fig.~\ref{spectra} and Table~\ref{fit_results}).
The 23.4-GHz maser was found to coincide spatially within the measurement uncertainty with the 
25-GHz masers, a well established class~I methanol maser series \citep[see, e.g.,][]{vor05}.
All these facts along with the detection of a rare class~I methanol maser at 9.9~GHz leave no
doubt that the 23.4-GHz maser we found towards G357.97$-$0.16 also belongs to the same class.
Moreover,  analogous to the 9.9 and bright 25-GHz masers, the 23.4-GHz masers are likely to originate from the areas with the most prominent interaction between shocks 
and quiescent medium, resulting in higher than average temperatures and densities. 
It is not clear at present whether 23.4~GHz masers
are as rare as the 9.9-GHz ones \citep[see][]{vor10} or are weak and common as the 25-GHz 
masers \citep[see][]{vor07}. Note, that the sensitivity and the spectral resolution attained in HOPS
allowed us to detect the strongest masers only. The survey sensitivity was suited primarily to
H$_2$O masers which often have flux densities a few orders of magnitude higher than the 
weak methanol masers.
Therefore, the single HOPS  detection at 23.4~GHz towards G357.97$-$0.16 is not surprising.
However, given the typical 5-$\sigma$ noise level and the spectral resolution of HOPS, it seems 
unlikely that there are many 23.4~GHz masers with a velocity integrated flux density in excess of
4~Jy~\kss (or peak flux density over 10~Jy assuming the same line width as in G357.97$-$0.16).

Interestingly, the relative flux densities of the 25-GHz maser series have 
the same trend with respect to the quantum number J as in G343.12$-$0.06 where a number of rare 
class~I methanol masers originate at single location associated with the strong 
shock traced by H$_2$ 2.12-$\mu$m emission 
\citep[c.f. Table~\protect\ref{fit_results} in this paper and Fig.~5 in][]{vor06}. In both sources, the maser in the J=2 transition is relatively weak while masers corresponding to higher J all have comparable
flux densities. In addition, both sources have rare 9.9-GHz masers. These similarities encouraged
us to search for 23.4-GHz masers towards G343.12$-$0.06. We examined CABB commissioning
data on the source (observed on 2010 June 3, archive project code CX110) and found a 23.4-GHz 
maser as a single spectral channel spike in the broad band spectrum which has a velocity resolution of 
about 13~\ks. The corresponding velocity integrated flux density was 1.6$\pm$0.2~Jy~\ks, which is
comparable to the value observed for the 25-GHz masers in the same source 
\citep[both according to these low-resolution data and to more 
accurate measurement of][]{vor06}, and the 9.9-GHz maser is only slightly stronger. Without a 
zoom window positioned to observe this maser with high velocity resolution
we are unable to quantify its intrinsic peak flux density. However, assuming  a spectral profile
similar to the 25-GHz masers, the flux density of the 23.4-GHz maser in G343.12$-$0.06 is expected to be around 7~Jy.  

Comparison of the relative integrated flux densities (or peak values) of the 9.9, 23.4 and
25-GHz masers in G343.12$-$0.06 and G357.97$-$0.16 \citep[c.f. Table~\protect\ref{fit_results} in this paper and Table~2 in][]{vor06} suggests that there is no linear relation between them. For example, 
the 23.4 and 25-GHz masers have comparable strength in G343.12$-$0.06, while the former 
is notably stronger in G357.97$-$0.16. Therefore, the new class~I maser will provide good constraints 
for maser models and should be included in future modelling. Further targeted observations,
which can be done routinely with a sub-Jy sensitivity at this frequency using ATCA, are required 
to advance our understanding of  the properties of 23.4-GHz masers. The obvious targets are the known
9.9 and 25-GHz masers.

\section{Conclusions}
\begin{enumerate}
\item We report the detection of a new class~I methanol maser in the $10_1-9_2$~A$^-$ transition
at 23.4~GHz. The only HOPS detection of such a maser was towards G357.97$-$0.16. It
was followed up at high angular resolution with ATCA 
along with the J$_2-$J$_1$~E class~I methanol maser series. Retrospectively, the 23.4~GHz
maser was also found towards G343.12$-$0.06, another prominent example where 
rare methanol masers are caused by strong shocks.
\item The absolute positions of the 23.4~GHz maser and the 25-GHz maser series coincide within
the measurement uncertainty in G357.97$-$0.16. These masers are located roughly on the 
extension of the line connecting 
two known sites of water and 6.7-GHz methanol maser emission, a few arcseconds offset from the southern site. We have detected a weak (1.3~mJy) continuum source associated with the northern 
site (famous for its water maser with an unusually wide velocity spread). This continuum
source may be a hyper-compact H{\sc ii} region.
\item We report the detection of the 9.9-GHz methanol maser in G357.97$-$0.16. It is only the
fifth maser found in this transition and also has the highest flux density (about 70~Jy) exceeding 
that of other known 9.9-GHz masers by an order of magnitude.
\item High spectral resolution measurement leads us to suggest that the rest frequencies for 
the J=2, 3 and 4
transitions of the J$_2-$J$_1$~E methanol series should be refined to 24934.379$\pm$0.001,
24928.7004$\pm$0.0008 and 24933.4698$\pm$0.0008~MHz, respectively.

\end{enumerate}

\section*{Acknowledgments}
The Australia Telescope Compact Array and the Mopra telescope are
parts of the Australia Telescope National Facility which is funded by the
Commonwealth of Australia for operation as a National Facility managed
by CSIRO.  The University of New South Wales Digital Filter Bank used for the 
observations with the Mopra Telescope was provided with support from the 
Australian Research Council. The research has made use of the NASA/IPAC Infrared
Science Archive, which is operated by the Jet Propulsion Laboratory,
California Institute of Technology, under contract with the National
Aeronautics and Space Administration.

\bsp

\label{lastpage}
\clearpage

\end{document}